\documentclass[11pt]{article}

\usepackage[margin=1in]{geometry}
\usepackage{amsmath,amssymb,amsthm}
\usepackage{graphicx}
\usepackage{booktabs}
\usepackage{caption}
\usepackage{subcaption}
\usepackage{cite}
\usepackage{xcolor}
\usepackage{indentfirst}
\usepackage{multirow}
\usepackage{float}
\usepackage{pgfplots}
\usepackage{pgfplotstable}
\usepackage{tikz}
\pgfplotsset{compat=1.14}
\usepackage{hyperref}

\definecolor{facered}{HTML}{FFBBBB}
\definecolor{facegreen}{HTML}{BBFFBB}
\definecolor{faceblue}{HTML}{BBDDFF}
\definecolor{edgered}{HTML}{CC0000}
\definecolor{edgegreen}{HTML}{008800}
\definecolor{edgeblue}{HTML}{0055CC}

\theoremstyle{definition}
\newtheorem{definition}{Definition}
\newtheorem{theorem}{Theorem}
\newtheorem{proposition}{Proposition}

\newtheorem{example}{Example}
\newtheorem{remark}{Remark}

\newcommand{\Stab}{\mathcal{S}}  % Stabilizer group
\newcommand{\stab}{S}            % Individual stabilizer operator
\newcommand{\Pauli}{\mathcal{P}}
\newcommand{\ket}[1]{|#1\rangle}

\newcommand{\demb}{d_{\text{emb}}}

\title{Hyperbolic and Semi-Hyperbolic Floquet Codes for Photonic Quantum Computing}
\author{Aygul Azatovna Galimova\textsuperscript{1}\\[4pt]
\textsuperscript{1}\textit{Department of Mathematics}\\
\textit{Duke University, Durham, NC 27708, USA}}
\date{}

\begin{document}

\maketitle

\begin{abstract}
Hyperbolic Floquet codes use only weight-2 measurements and can be implemented directly on hardware with native pair measurements. We construct hyperbolic and semi-hyperbolic Floquet codes from $\{8,3\}$, $\{10,3\}$, and $\{12,3\}$ tessellations via the Wythoff kaleidoscopic construction with the Low-Index Normal Subgroups (LINS) algorithm. The $\{10,3\}$ and $\{12,3\}$ families are new to hyperbolic Floquet codes. We evaluate these codes under four noise models. Under ancilla-based Entangling Measurement (EM3) noise, all three families achieve a threshold of ${\sim}1.5\%$. With a native pair-measurement depolarizing model (SDEM3), thresholds are ${\sim}1.0$--$1.2\%$. For heralded photon loss, the $\{8,3\}$ family achieves ${\sim}8.5$--$9\%$, exceeding the planar honeycomb threshold of ${\sim}6.3\%$. In the multi-parameter SPOQC-2 noise model, the $\{8,3\}$ codes achieve a 2D fault-tolerant area $2.2\times$ that of the surface code compiled to pair measurements. We present the first photon loss and SPOQC-2 thresholds for hyperbolic Floquet codes.
\end{abstract}

\tableofcontents
\newpage

%==============================================================================
\section{Introduction}
%==============================================================================

\indent Quantum error correction is necessary for fault-tolerant quantum computation~\cite{nielsen2010quantum, shor1995scheme}. Topological codes such as the surface code~\cite{kitaev2003fault, dennis2002topological} are widely studied for near-term quantum error correction due to their planar geometry and nearest-neighbor connectivity~\cite{fowler2012surface}.

\indent A limitation of planar surface codes is their encoding rate: the ratio $k/n$ of logical to physical qubits approaches zero as code size increases~\cite{bravyi2010tradeoffs}. Hyperbolic surface codes achieve constant rate $k/n = \Theta(1)$ by embedding qubits on negatively curved surfaces~\cite{breuckmann2016constructions}. However, their face stabilizers have weight $p$ (e.g., weight 8 for $\{8,3\}$) and require multi-qubit parity measurements~\cite{breuckmann2016constructions}.

\indent Floquet codes~\cite{hastings2021dynamically} resolve this problem by replacing high-weight stabilizers with periodic sequences of weight-2 measurements that dynamically generate the code space. Fahimniya et al.~\cite{fahimniya2023faulttolerant} apply this idea to $\{8,3\}$ hyperbolic tessellations, producing Floquet codes with constant encoding rate and weight-2 measurements. Higgott and Breuckmann~\cite{higgott2024constructions} extend the construction to semi-hyperbolic (fine-grained) variants with improved distance scaling ($\demb = O(\sqrt{n})$) and show that the resulting codes are over 100$\times$ more efficient than surface codes when compiled to pair measurements at physical error rate $p = 0.1\%$. Ozawa et al.~\cite{ozawa2025hyperbolic} propose a simplified schedule using only $XX$ and $ZZ$ measurements.

\indent The restriction to weight-2 measurements makes these codes natural candidates for hardware with native pair measurements. On most platforms, implementing a two-qubit parity measurement requires an ancilla circuit (reset, two CNOTs, measure), incurring idle decoherence and gate errors. Two emerging architectures avoid this overhead. Majorana tetrons perform parity measurements directly, without ancilla qubits~\cite{paetznick2023performance}. In the spin-optical quantum computing (SPOQC) architecture~\cite{dessertaine2024enhanced}, parity measurements are implemented through photon-mediated entanglement, and every photon loss event is heralded and induces a phase erasure channel on the spin qubits. These two platforms motivate distinct noise models: depolarizing (SDEM3) for Majorana tetrons and photon loss (modeled as erasure) for spin-optical qubits. On the planar honeycomb code, Dessertaine et al.~\cite{dessertaine2024enhanced} demonstrate a photon loss threshold of ${\sim}6.3\%$ under this erasure model. No prior work has evaluated hyperbolic Floquet codes under photon loss noise or explored tessellation families beyond $\{8,3\}$.

\indent We construct hyperbolic and semi-hyperbolic Floquet codes from $\{8,3\}$, $\{10,3\}$, and $\{12,3\}$ tessellations using the LINS algorithm with the Wythoff kaleidoscopic construction and evaluate them under the following noise models:
\begin{enumerate}
\item Phenomenological noise (Section~\ref{sec:phenomenological}): independent data and measurement errors, used for comparison with prior work.
\item Correlated EM3 noise (Section~\ref{sec:circuit_level}): ancilla-based measurement circuits with correlated Pauli and flip errors.
\item SDEM3 (Single-step Depolarizing EM3) noise (Section~\ref{sec:sdem3}): depolarizing model motivated by Majorana tetron architectures.
\item Photon loss noise (Section~\ref{sec:erasure}): photon loss on spin-optical links, heralded and modeled as erasure.
\end{enumerate}
In a companion paper~\cite{galimova2026distributed}, we evaluate the same code families in a distributed setting.

%==============================================================================
\section{Stabilizer Codes and Floquet Codes}
%==============================================================================

\subsection{The Stabilizer Formalism}

\indent The stabilizer formalism~\cite{gottesman1997stabilizer} provides a group-theoretic framework for quantum error correction. For $n$ qubits, the Pauli group is $\Pauli_n = \{\pm 1, \pm i\} \times \{I, X, Y, Z\}^{\otimes n}$.

\begin{definition}[Stabilizer Code]
An $[[n, k, d]]$ stabilizer code is a $2^k$-dimensional subspace of $(\mathbb{C}^2)^{\otimes n}$ determined by an abelian subgroup $\Stab \subseteq \Pauli_n$:
\begin{equation}
\mathcal{C} = \{\ket{\psi} \in (\mathbb{C}^2)^{\otimes n} : g\ket{\psi} = \ket{\psi} \text{ for all } g \in \Stab\},
\end{equation}
where $\Stab$ does not contain $-I$, $|\Stab| = 2^{n-k}$, and $\Stab = \langle g_1, \ldots, g_{n-k} \rangle$.
\end{definition}

\begin{definition}[Logical Operators]
Logical operators are elements of the normalizer modulo the stabilizer:
\begin{equation}
\text{Logicals} = N(\Stab) / \Stab,
\end{equation}
where $N(\Stab) = \{P \in \Pauli_n : PgP^{-1} \in \Stab \text{ for all } g \in \Stab\}$.
\end{definition}

\begin{definition}[Code Distance]
The distance of a stabilizer code is:
\begin{equation}
d = \min\{\text{wt}(L) : L \in N(\Stab) \setminus \Stab\},
\end{equation}
where $\text{wt}(L)$ is the number of non-identity tensor factors of the Pauli operator $L$.
\end{definition}

\indent For Floquet codes, the code space is dynamically generated, and the relevant distance metric is the \emph{embedded distance} $\demb$~\cite{higgott2024constructions}, defined in Section~\ref{sec:code_construction}.

\begin{definition}[CSS Code]
A Calderbank-Shor-Steane (CSS) code is a stabilizer code whose generators partition into $X$-type (products of $X$ only) and $Z$-type (products of $Z$ only).
\end{definition}

\subsection{Surface Code Stabilizers}

\indent For CSS surface codes on tessellations, qubits are placed on edges with stabilizers associated with faces and vertices. We write $\Stab$ for the stabilizer group and use $\stab_f$, $\stab_v$ for individual stabilizer operators (elements of $\Stab$).

\begin{definition}[Plaquette and Vertex Stabilizers]
For each face $f$, define $\stab_f^Z = \prod_{e \in \partial f} Z_e$. For each vertex $v$, define $\stab_v^X = \prod_{e \in \text{star}(v)} X_e$.
\end{definition}

\indent For hyperbolic Floquet codes, the qubit placement differs: physical qubits are placed on vertices, and two-body measurements act on edges (pairs of adjacent vertices). Each edge connects exactly two vertices, so $XX$, $YY$, or $ZZ$ measurements act between adjacent qubits.

\begin{proposition}[Number of Logical Qubits]
For a surface code on a genus-$g$ surface with $V$ vertices, $E$ edges, and $F$ faces:
\begin{equation}
k = E - (F-1) - (V-1) = E - F - V + 2 = 2g,
\end{equation}
using the Euler characteristic $\chi = V - E + F = 2 - 2g$. The $F$ face stabilizers satisfy one linear dependency ($\prod_f \stab_f^Z = I$) and therefore have rank $F - 1$. The $V$ vertex stabilizers satisfy one dependency ($\prod_v \stab_v^X = I$) and have rank $V - 1$~\cite{breuckmann2016constructions}.
\end{proposition}

\subsection{The Floquet Measurement Cycle}

\indent A Floquet code~\cite{hastings2021dynamically} is defined by a periodic sequence of non-commuting two-body measurements that dynamically generate a code space~\cite{fahimniya2023faulttolerant}. In each round, a set of commuting weight-2 Pauli operators is measured; operators in consecutive rounds generally anti-commute. Our $\{p,3\}$ tessellations are 3-edge-colorable (Section~\ref{sec:face_coloring}), so we adopt the $XX/YY/ZZ$ schedule~\cite{fahimniya2023faulttolerant, higgott2024constructions}, which cycles through three rounds:
\begin{itemize}
\item Round 0 (Red): Measure $XX$ on all red edges.
\item Round 1 (Green): Measure $YY$ on all green edges.
\item Round 2 (Blue): Measure $ZZ$ on all blue edges.
\end{itemize}

\indent The \emph{instantaneous stabilizer group} (ISG) is the stabilizer group of the code at a given point in the schedule~\cite{hastings2021dynamically}. Each round updates the ISG: the new check measurements replace anti-commuting generators from the previous round. After reaching steady state, the state is in the $+1$-eigenspace of all plaquette stabilizers~\cite{higgott2024constructions}.

\begin{definition}[Floquet Face Stabilizer]
\label{def:floquet_face}
The face stabilizer $\stab_f$ for a $p$-gon face $f$ is the product of the $p$ edge checks around the boundary:
\begin{equation}
\stab_f = \prod_{e \in \partial f} M_e,
\end{equation}
where $M_e \in \{XX, YY, ZZ\}$ depends on the edge color.
\end{definition}

\subsection{Weight-2 Decomposition}

\indent High-weight face stabilizers can be inferred from products of weight-2 edge measurements. Consider a hexagonal face with alternating edge colors R, G, B, R, G, B. Writing $M_{e_i}$ for the measurement on edge $e_i$ as in Definition~\ref{def:floquet_face}, the face stabilizer is:
\begin{equation}
\stab_f = M_{e_1} \cdot M_{e_2} \cdot M_{e_3} \cdot M_{e_4} \cdot M_{e_5} \cdot M_{e_6} = (XX)_1 \cdot (YY)_2 \cdot (ZZ)_3 \cdot (XX)_4 \cdot (YY)_5 \cdot (ZZ)_6,
\end{equation}
where indices label edges around the face. Each vertex is incident to exactly two edges with different colors. Two different Paulis therefore act on each vertex. The product simplifies using $XY = iZ$, $YZ = iX$, $ZX = iY$. Detectors are constructed by comparing plaquette measurement outcomes across consecutive cycles~\cite{higgott2024constructions}.

%==============================================================================
\section{Hyperbolic Geometry and Tessellations}
\label{sec:hyperbolic_geometry}
%==============================================================================

\subsection{The Poincar\'e Disk Model}

\indent The hyperbolic plane $\mathbb{H}^2$ is a two-dimensional Riemannian manifold with constant negative Gaussian curvature $K = -1$. In the Poincar\'e disk model $\mathbb{D} = \{z \in \mathbb{C} : |z| < 1\}$, the metric is:
\begin{equation}
ds^2 = \frac{4|dz|^2}{(1-|z|^2)^2}.
\end{equation}

\indent The distance between two points $z_1, z_2 \in \mathbb{D}$ is:
\begin{equation}
\text{dist}(z_1, z_2) = 2 \text{arctanh} \left( \frac{|z_1 - z_2|}{|1 - z_1 \bar{z}_2|} \right).
\end{equation}
The isometries of $\mathbb{D}$ are M\"obius transformations in $\text{PSU}(1,1)$. These transformations preserve hyperbolic distances but distort Euclidean lengths.

\subsection{Fuchsian Groups and Quotient Surfaces}

\indent A Fuchsian group is a discrete subgroup of $\text{PSL}(2, \mathbb{R})$, the group of orientation-preserving isometries of $\mathbb{H}^2$. Each Fuchsian group $\Gamma$ has a fundamental domain $\mathcal{F} \subset \mathbb{H}^2$ such that translates of $\mathcal{F}$ tile the hyperbolic plane.

\begin{definition}[Fundamental Domain]
A fundamental domain for a Fuchsian group $\Gamma$ is a connected region $\mathcal{F} \subset \mathbb{H}^2$ such that:
\begin{enumerate}
\item $\bigcup_{\gamma \in \Gamma} \gamma(\mathcal{F}) = \mathbb{H}^2$ (the translates cover the plane),
\item $\gamma_1(\mathcal{F})^\circ \cap \gamma_2(\mathcal{F})^\circ = \emptyset$ for $\gamma_1 \neq \gamma_2$ (interiors are disjoint).
\end{enumerate}
\end{definition}

\indent When $\Gamma$ is torsion-free (no elements of finite order except identity), the quotient $\mathbb{H}^2 / \Gamma$ is a closed orientable surface of genus $g \geq 2$. The genus is determined by the hyperbolic area of the fundamental domain via the Gauss-Bonnet theorem:
\begin{equation}
\text{Area}(\mathcal{F}) = 2\pi(2g - 2) = -2\pi\chi,
\end{equation}
where $\chi = 2 - 2g$ is the Euler characteristic.

\subsection{Schl\"afli Symbols and Tessellations}

\begin{definition}[Schl\"afli Symbol]
A $\{p, q\}$ tessellation is a regular tiling where each face is a regular $p$-gon and each vertex has degree $q$.
\end{definition}

\begin{theorem}[Hyperbolic Condition]
A $\{p,q\}$ tessellation exists in the hyperbolic plane if and only if:
\begin{equation}
(p-2)(q-2) > 4.
\end{equation}
Equivalently: $\frac{1}{p} + \frac{1}{q} < \frac{1}{2}$.
\end{theorem}

\indent For our code families:
\begin{itemize}
\item $\{8,3\}$: $(8-2)(3-2) = 6 > 4$.
\item $\{10,3\}$: $(10-2)(3-2) = 8 > 4$.
\item $\{12,3\}$: $(12-2)(3-2) = 10 > 4$.
\end{itemize}
Figures~\ref{fig:tess_8_3}--\ref{fig:tess_12_3} illustrate these three tessellations in the Poincar\'e disk model.

\begin{figure}[htbp]
\centering

\caption{The $\{8,3\}$ tessellation of the hyperbolic plane in the Poincar\'e disk model. Each face is a regular octagon with three faces meeting at every vertex. Edges are 3-colored (red, green, blue), guaranteeing a valid Floquet measurement schedule.}
\label{fig:tess_8_3}
\end{figure}

\begin{figure}[htbp]
\centering
%
\caption{The $\{10,3\}$ tessellation. Each face is a regular decagon. The larger face size yields a higher edge-to-vertex ratio.}
\label{fig:tess_10_3}
\end{figure}

\begin{figure}[htbp]
\centering
%
\caption{The $\{12,3\}$ tessellation. Each face is a regular dodecagon. The exponential shrinking of faces toward the disk boundary reflects the negative curvature of the hyperbolic plane.}
\label{fig:tess_12_3}
\end{figure}

\subsection{Quotient Surfaces and Combinatorial Parameters}

\indent A closed hyperbolic surface $\Sigma_g$ of genus $g \geq 2$ is realized as a quotient $\mathbb{H}^2 / \Gamma$, where $\Gamma$ is a discrete, torsion-free Fuchsian group. The Euler characteristic relates the cell counts:
\begin{equation}
\chi = V - E + F = 2 - 2g.
\end{equation}

\begin{proposition}[Combinatorial Parameters]
For a uniform $\{p,q\}$ tessellation of a genus-$g$ surface (all faces are $p$-gons, all vertices have degree $q$), the constraints $qV = 2E = pF$ yield:
\begin{equation}
F = \frac{4q(g-1)}{pq - 2p - 2q}, \quad E = \frac{pF}{2}, \quad V = \frac{pF}{q}.
\end{equation}
This applies to base codes; fine-grained codes have subdivision vertices with degree up to 6.
\end{proposition}

\begin{example}[Computing Parameters for $\{8,3\}$, Genus 2]
For $\{8,3\}$ on genus-2:
\begin{align}
F &= \frac{4 \cdot 3 \cdot (2-1)}{8 \cdot 3 - 2 \cdot 8 - 2 \cdot 3} = \frac{12}{24 - 16 - 6} = \frac{12}{2} = 6, \\
E &= \frac{8 \cdot 6}{2} = 24, \quad V = \frac{8 \cdot 6}{3} = 16.
\end{align}
Verification: $V - E + F = 16 - 24 + 6 = -2 = 2 - 2(2)$. \checkmark
\end{example}

\subsection{Asymptotic Scaling}

\indent The distinction between Euclidean and hyperbolic geometry affects quantum code parameters. In Euclidean geometry, area grows polynomially with radius ($A \sim r^2$), while in hyperbolic geometry, area grows exponentially ($A \sim e^r$).

\begin{proposition}[Euclidean Scaling]
For surface codes on Euclidean tessellations (e.g., the square lattice $\{4,4\}$):
\begin{equation}
n = \Theta(d^2), \quad k = \Theta(1), \quad \frac{k}{n} = \Theta(1/n) \to 0.
\end{equation}
The encoding rate vanishes as code size increases.
\end{proposition}

\begin{theorem}[Hyperbolic Scaling]
For hyperbolic surface codes with $n$ physical qubits~\cite{breuckmann2016constructions}:
\begin{equation}
\frac{k}{n} = \Theta(1) \quad \text{(constant rate)}, \quad \demb = \Theta(\log n) \quad \text{(logarithmic distance)}.
\end{equation}
For edge-based codes on $\{p,q\}$ tessellations, the asymptotic rate is $1 - 2/p - 2/q$~\cite{breuckmann2016constructions}. This rate-distance tradeoff is analyzed in~\cite{breuckmann2016constructions}. Fine-graining (Section~\ref{sec:fine_graining}) improves the distance scaling to $\demb = O(\sqrt{n})$ at fixed $k$. The resulting codes are termed semi-hyperbolic.
\end{theorem}

\indent Hyperbolic tessellations provide the favorable scaling properties, but realizing them as Floquet codes requires an additional structure: a 3-coloring of faces such that no two faces sharing an edge have the same color. This 3-coloring determines the edge partition (red, green, blue) that defines the Floquet measurement schedule. Not all hyperbolic tessellations admit such a coloring, and verifying 3-colorability post hoc requires a separate computation. The Wythoff kaleidoscopic construction avoids this problem, as described in the next section.

%==============================================================================
\section{Wythoff Construction}
\label{sec:face_coloring}
%==============================================================================

\indent Color code tilings on the hyperbolic plane can be constructed through Wythoff's kaleidoscopic method~\cite{higgott2024constructions}. A generator point is placed inside the fundamental triangle and perpendicular lines are projected from that point to each of the bounding mirrors. The resulting tilings have 3-colorable faces. The tiling is generated by a triangle group; quotienting by a normal subgroup yields a finite tiling on a closed surface. Each edge inherits its color as the complement of its two adjacent face colors:
\begin{equation}
\text{color}(e) = 3 - \text{color}(f_1) - \text{color}(f_2),
\end{equation}
where $f_1, f_2$ are the faces incident to edge $e$.

\subsection{Triangle Groups}

\begin{definition}[Triangle Group]
The triangle group $\Delta(p,q,r)$ has presentation:
\begin{equation}
\Delta(p,q,r) = \langle a, b, c \mid a^2 = b^2 = c^2 = (ab)^r = (bc)^q = (ca)^p = e \rangle,
\end{equation}
where $a, b, c$ are reflections across the three sides. The fundamental triangle has interior angles $\pi/p$, $\pi/q$, $\pi/r$ at its vertices.
\end{definition}

\indent The sum of interior angles determines the geometry:
\begin{itemize}
\item $\frac{1}{p} + \frac{1}{q} + \frac{1}{r} > 1$: spherical geometry.
\item $\frac{1}{p} + \frac{1}{q} + \frac{1}{r} = 1$: Euclidean geometry.
\item $\frac{1}{p} + \frac{1}{q} + \frac{1}{r} < 1$: hyperbolic geometry.
\end{itemize}

\indent For a $\{p,q\}$ tessellation, we use $\Delta(p,q,2)$ with $r=2$:
\begin{equation}
\Delta(p,q,2) = \langle a, b, c \mid a^2 = b^2 = c^2 = (ab)^2 = (bc)^q = (ca)^p = e \rangle.
\end{equation}
The group acts on the hyperbolic plane when $\frac{1}{p} + \frac{1}{q} + \frac{1}{2} < 1$. For our families with $q = 3$:
\begin{itemize}
\item $\{8,3\}$: $\frac{1}{8} + \frac{1}{3} + \frac{1}{2} = \frac{23}{24} < 1$.
\item $\{10,3\}$: $\frac{1}{10} + \frac{1}{3} + \frac{1}{2} = \frac{14}{15} < 1$.
\item $\{12,3\}$: $\frac{1}{12} + \frac{1}{3} + \frac{1}{2} = \frac{11}{12} < 1$.
\end{itemize}

\begin{remark}[Rotation Subgroup]
For computation, we use the rotation subgroup $\Delta^+(p,q,2) = \langle \alpha, \beta, \gamma \mid \alpha^2 = \beta^q = \gamma^p = \alpha\beta\gamma = e \rangle$. This index-2 subgroup consists of orientation-preserving isometries. The correspondence with the reflection group is $\alpha = ab$, $\beta = bc$, $\gamma = ca$. Both presentations yield the same quotient tessellations.
\end{remark}

\subsection{Flag Structure and Coset Enumeration}

\indent Our pipeline extracts the tessellation combinatorics from the coset table using the flag structure of the tiling. A flag in the tessellation is an incident triple (vertex, edge, face). The triangle group acts transitively on flags with trivial stabilizer: each coset of the identity corresponds to exactly one flag.

\begin{definition}[Coset Table]
Given a finite-index normal subgroup $N \trianglelefteq \Delta(p,q,2)$, the coset table records the action of generators on cosets. The number of cosets equals the number of flags in the quotient tessellation.
\end{definition}

\indent The tessellation structure emerges from orbits under subgroups of the rotation group $\Delta^+$:
\begin{itemize}
\item Vertices: orbits under $\langle \beta \rangle$ (each orbit has $q$ elements).
\item Edges: orbits under $\langle \alpha \rangle$ (each orbit has 2 elements).
\item Faces: orbits under $\langle \gamma \rangle$ (each orbit has $p$ elements).
\end{itemize}

%==============================================================================
\section{Code Construction}
\label{sec:code_construction}
%==============================================================================

\indent Section~\ref{sec:hyperbolic_geometry} established that hyperbolic Floquet codes arise from tessellations of closed hyperbolic surfaces, with the Wythoff construction providing the geometric framework: triangle groups define tessellations, and quotients by normal subgroups yield closed surfaces with 3-colorable faces. The Wythoff construction tells us \emph{how} to build a tessellation from a given subgroup, but does not enumerate which subgroups exist.

\indent To enumerate all such quotients systematically, we use the Low-Index Normal Subgroups (LINS) algorithm. The coset table from LINS encodes exactly the flag structure needed to extract vertices, edges, faces, and their incidences. This section describes the enumeration pipeline and the resulting code parameters.

\subsection{The LINS Algorithm}

\indent The LINS algorithm enumerates finite-index normal subgroups of finitely presented groups. We use the GAP~\cite{gap4} implementation provided by the LINS package~\cite{lins, firth2005} to enumerate subgroups of the rotation group $\Delta^+(p,3,2)$. The coset table for $N$ encodes the tessellation combinatorics: vertices are orbits under the generator of order 3, edges are orbits under the generator of order 2, and faces are orbits under the generator of order $p$. Table~\ref{tab:pipeline_comparison} compares our construction pipeline with prior work.

\subsection{Comparison with Prior Construction Methods}

\begin{table}[htbp]
\centering
\caption{Comparison of hyperbolic code construction methods.}
\label{tab:pipeline_comparison}
\resizebox{\columnwidth}{!}{%
\begin{tabular}{@{}lcccccc@{}}
\toprule
 & Fahimniya et al. & Higgott et al. & Ozawa et al. & Mahmoud et al. & Sutcliffe et al. & This work \\
\midrule
Tessellations & $\{8,3\}$ & $\{8,3\}$ & $\{8,3\}$ & $\{8,3\}$, $\{10,3\}$ & $\{8,3\}$ & $\{p,3\}$\rlap{$^*$} \\
Code type & Floquet & Floquet & Floquet (HCF) & Static CSS & Floquet & Floquet \\
Measurement schedule & XX/YY/ZZ & XX/YY/ZZ & XX/ZZ & --- & XX/YY/ZZ & XX/YY/ZZ \\
Construction & Conder & Wythoff & Fahimniya+Higgott & LINS+Bravais & Foster+Higgott & LINS+Wythoff \\
\bottomrule
\multicolumn{7}{@{}l}{\scriptsize $^*$\,Polygon order $p \in \{8,10,12\}$.}
\end{tabular}}
\end{table}

\indent Fahimniya et al.~\cite{fahimniya2023faulttolerant} select $\{8,3\}$ lattices from the Conder database~\cite{conder2002trivalent} of symmetric trivalent graphs and filter by girth~8 and valid face 3-colorings. Higgott and Breuckmann~\cite{higgott2024constructions} use the Wythoff construction and provide Stim circuit files but do not publish generator code. Ozawa et al.~\cite{ozawa2025hyperbolic} apply a six-step XX/ZZ-only schedule to $\{8,3\}$ hyperbolic Floquet codes, citing Fahimniya et al.\ and Higgott and Breuckmann for the underlying code construction. Mahmoud et al.~\cite{mahmoud2025systematic} construct hyperbolic surface codes on $\{p,q\}$ tessellations with specific underlying Bravais lattices ($\{4g,4g\}$ and $\{2(2g{+}1),2g{+}1\}$); they simulate $\{8,3\}$ and $\{10,3\}$ hyperbolic surface codes. Sutcliffe et al.~\cite{sutcliffe2025distributed} take $\{8,3\}$ graphs from the Foster census~\cite{foster1988census, foster2020encyclopedia} and adopt the Floquet code construction of Higgott and Breuckmann~\cite{higgott2024constructions}; they evaluate these codes in distributed settings but do not introduce new code families. Our pipeline derives the tessellation structure directly from the Wythoff construction and uses LINS for subgroup enumeration. The result is a parametric generator for arbitrary $\{p,3\}$ families.

\indent Applying this pipeline to $\{8,3\}$, $\{10,3\}$, and $\{12,3\}$ tessellations produces the Floquet code families described below. We use the same pipeline in our distributed hyperbolic Floquet and hyperbolic CSS erasure-threshold papers~\cite{galimova2026distributed,galimova2026erasure}.

\subsection{Code Parameters}

\indent Table~\ref{tab:code_params} summarizes the codes constructed. For each code, $n = V$ is the number of physical qubits (vertices), $E$ the number of edges, $F$ the number of faces, $g$ the genus, $k = 2g$ the number of logical qubits, and $\demb$ is the embedded distance~\cite{higgott2024constructions}. For each color $c \in \{R,G,B\}$, the restricted dual lattice $T^*_c$ is the dual graph with faces of color $c$ removed; $\demb$ equals the minimum weight of a homologically non-trivial cycle or co-cycle in any $T^*_c$, measuring the minimum number of single-qubit errors that can cause a logical failure under the Floquet decoding scheme~\cite{higgott2024constructions}.

\begin{table}[htbp]
\centering
\caption{Code parameters for hyperbolic Floquet codes constructed in this work. Codes marked $^\dagger$ have $\demb = 2$ and cannot correct errors; they are excluded from threshold estimation but retained as base codes for fine-graining (Section~\ref{sec:fine_graining}). All $\{12,3\}$ base codes have $\demb = 2$ (no $^\dagger$ markers needed).}
\label{tab:code_params}
\begin{tabular}{llrrrrrrrr}
\toprule
Family & Code & $n$ & $E$ & $F$ & $g$ & $k$ & $k/n$ & $\demb$ & Source \\
\midrule
\multirow{7}{*}{$\{8,3\}$}
& H16$_{8,3}^\dagger$ & 16 & 24 & 6 & 2 & 4 & 0.25 & 2 & \cite{fahimniya2023faulttolerant,higgott2024constructions} \\
& H32$_{8,3}^\dagger$ & 32 & 48 & 12 & 3 & 6 & 0.19 & 2 & \cite{fahimniya2023faulttolerant} \\
& H64$_{8,3}^\dagger$ & 64 & 96 & 24 & 5 & 10 & 0.16 & 2 & \cite{fahimniya2023faulttolerant,higgott2024constructions} \\
& H144$_{8,3}$ & 144 & 216 & 54 & 10 & 20 & 0.14 & 4 & \cite{fahimniya2023faulttolerant,higgott2024constructions} \\
& H256$_{8,3}$ & 256 & 384 & 96 & 17 & 34 & 0.13 & 4 & This work \\
& H336$_{8,3}$ & 336 & 504 & 126 & 22 & 44 & 0.13 & 4 & This work \\
& H432$_{8,3}$ & 432 & 648 & 162 & 28 & 56 & 0.13 & 4 & This work \\
\midrule
\multirow{4}{*}{$\{10,3\}$}
& H50$_{10,3}^\dagger$ & 50 & 75 & 15 & 6 & 12 & 0.24 & 2 & This work \\
& H120$_{10,3}^\dagger$ & 120 & 180 & 36 & 13 & 26 & 0.22 & 2 & This work \\
& H250$_{10,3}$ & 250 & 375 & 75 & 26 & 52 & 0.21 & 4 & This work \\
& H720$_{10,3}$ & 720 & 1080 & 216 & 73 & 146 & 0.20 & 4 & This work \\
\midrule
\multirow{5}{*}{$\{12,3\}$}
& H48$_{12,3}$ & 48 & 72 & 12 & 7 & 14 & 0.29 & 2 & This work \\
& H72$_{12,3}$ & 72 & 108 & 18 & 10 & 20 & 0.28 & 2 & This work \\
& H96$_{12,3}$ & 96 & 144 & 24 & 13 & 26 & 0.27 & 2 & This work \\
& H168$_{12,3}$ & 168 & 252 & 42 & 22 & 44 & 0.26 & 2 & This work \\
& H312$_{12,3}$ & 312 & 468 & 78 & 40 & 80 & 0.26 & 2 & This work \\
\bottomrule
\end{tabular}
\end{table}

\indent All $\{12,3\}$ base codes have $\demb = 2$. In contrast, $\{8,3\}$ codes achieve $\demb \geq 4$ only at larger code sizes (H144 and above), and $\{10,3\}$ codes require even larger sizes (H250 and above). The $\{12,3\}$ family therefore depends entirely on fine-graining (Section~\ref{sec:fine_graining}) to achieve error-correcting distance.

\subsection{Fine-Graining and Semi-Hyperbolic Codes}
\label{sec:fine_graining}

\subsubsection{The Fine-Graining Procedure}

\indent Hyperbolic Floquet codes achieve constant encoding rate $k/n = \Theta(1)$ but have distance $\demb = O(\log n)$. This logarithmic scaling limits error suppression at fixed code size. Breuckmann et al.~\cite{breuckmann2017semihyperbolic} introduced a fine-graining procedure for hyperbolic CSS surface codes that increases distance while preserving the number of logical qubits. Higgott and Breuckmann~\cite{higgott2024constructions} extended this procedure to hyperbolic Floquet codes. The resulting codes are termed semi-hyperbolic because they interpolate between hyperbolic ($\demb = O(\log n)$) and Euclidean ($\demb = O(\sqrt{n})$) distance scaling.

\indent The fine-graining procedure subdivides each triangle in the dual lattice into $\ell^2$ smaller triangles, where $\ell$ is the subdivision parameter. This is achieved by inserting $\ell-1$ vertices along each edge and $(\ell-1)(\ell-2)/2$ interior vertices per face. For a base code with parameters $[[n, k, \demb]]$, fine-graining produces a code with:
\begin{equation}
n' = \ell^2 \cdot n, \quad k' = k.
\end{equation}
This formula holds for vertex-based Floquet codes. The key observation is that physical qubits sit at primal vertices. Primal vertices correspond bijectively to dual faces (triangles). Subdividing each dual triangle into $\ell^2$ sub-triangles creates $\ell^2$ sub-faces per original face, hence $\ell^2$ times as many primal vertices. The shared dual vertices at triangle boundaries do not affect this count because we are counting faces (= primal vertices), not dual vertices. The genus and number of logical qubits remain unchanged because fine-graining is a topological refinement that preserves the surface structure.

\indent The embedded distance $\demb$ must be computed independently for each fine-grained code (Table~\ref{tab:fine_grained_codes}). Higgott and Breuckmann observe that semi-hyperbolic codes achieve $\demb = O(\sqrt{n})$~\cite{higgott2024constructions}. For $n' = \ell^2 n$, this gives $\demb = O(\ell)$.

\begin{definition}[Fine-Graining Level]
Level $\ell = 1$ is the base code with no subdivision. Level $\ell = 2$ subdivides each edge into 2 segments (each triangle into 4). Level $\ell = 3$ subdivides each edge into 3 segments (each triangle into 9).
\end{definition}

\subsubsection{Geometric Construction}

\indent For codes derived from hyperbolic tessellations, fine-graining respects the hyperbolic metric to maintain the 3-coloring required for the Floquet measurement schedule. New vertices along edges are placed at hyperbolic geodesic midpoints using M\"obius transformations in the Poincar\'e disk model.

\indent Given two points $z_1, z_2 \in \mathbb{D}$, the geodesic midpoint is computed by:
\begin{enumerate}
\item Map $z_1$ to the origin via the isometry $\phi(z) = (z - z_1)/(1 - \bar{z}_1 z)$.
\item Find the hyperbolic midpoint along the radial geodesic: $w = \phi(z_2)/(1 + \sqrt{1 - |\phi(z_2)|^2})$.
\item Map back via $\phi^{-1}(w) = (w + z_1)/(1 + \bar{z}_1 w)$.
\end{enumerate}
Interior vertices are placed using hyperbolic barycentric interpolation within each triangle.

\subsubsection{Color Inheritance}

\indent The 3-coloring of faces (equivalently, edges) extends consistently to the fine-grained lattice. The subdivision of each triangular cell in the dual lattice inherits colors from its three corner vertices. These corner vertices have distinct colors from the base tessellation's face 3-coloring. New vertices introduced by subdivision receive colors such that adjacent vertices always have different colors. This is achieved by assigning colors via linear interpolation in barycentric coordinates modulo 3. The resulting edge coloring partitions edges into three disjoint classes. This preserves the measurement schedule structure required for hyperbolic Floquet codes.

\subsubsection{Comparison: Hyperbolic vs Semi-Hyperbolic}

\begin{table}[htbp]
\centering
\caption{Base codes and fine-grained variants. Fine-graining increases $n$ and $\demb$ while preserving $k$.}
\label{tab:fine_grained_codes}
\begin{tabular}{llrrrrl}
\toprule
Base & Level & $n$ & $k$ & $\demb$ & $k/n$ & Type \\
\midrule
H16$_{8,3}$ & $\ell=1$ & 16 & 4 & 2 & 0.25 & Hyperbolic \\
H16$_{8,3}$ & $\ell=2$ & 64 & 4 & 3 & 0.06 & Semi-hyperbolic \\
H16$_{8,3}$ & $\ell=3$ & 144 & 4 & 4 & 0.03 & Semi-hyperbolic \\
H16$_{8,3}$ & $\ell=4$ & 256 & 4 & 6 & 0.02 & Semi-hyperbolic \\
H16$_{8,3}$ & $\ell=5$ & 400 & 4 & 7 & 0.01 & Semi-hyperbolic \\
\midrule
H64$_{8,3}$ & $\ell=1$ & 64 & 10 & 2 & 0.16 & Hyperbolic \\
H64$_{8,3}$ & $\ell=2$ & 256 & 10 & 4 & 0.04 & Semi-hyperbolic \\
H64$_{8,3}$ & $\ell=3$ & 576 & 10 & 6 & 0.02 & Semi-hyperbolic \\
H64$_{8,3}$ & $\ell=4$ & 1024 & 10 & 10 & 0.01 & Semi-hyperbolic \\
\midrule
H50$_{10,3}$ & $\ell=1$ & 50 & 12 & 2 & 0.24 & Hyperbolic \\
H50$_{10,3}$ & $\ell=2$ & 200 & 12 & 4 & 0.06 & Semi-hyperbolic \\
H50$_{10,3}$ & $\ell=3$ & 450 & 12 & 6 & 0.03 & Semi-hyperbolic \\
H50$_{10,3}$ & $\ell=4$ & 800 & 12 & 8 & 0.02 & Semi-hyperbolic \\
\midrule
H48$_{12,3}$ & $\ell=1$ & 48 & 14 & 2 & 0.29 & Hyperbolic \\
H48$_{12,3}$ & $\ell=2$ & 192 & 14 & 4 & 0.07 & Semi-hyperbolic \\
H48$_{12,3}$ & $\ell=3$ & 432 & 14 & 4 & 0.03 & Semi-hyperbolic \\
H48$_{12,3}$ & $\ell=4$ & 768 & 14 & 7 & 0.02 & Semi-hyperbolic \\
\bottomrule
\end{tabular}
\end{table}

\indent Table~\ref{tab:fine_grained_codes} compares base codes with their fine-grained variants. Fine-graining improves distance at the cost of encoding rate. A base H16$_{8,3}$ code with $k/n = 0.25$ becomes a semi-hyperbolic code with $k/n = 0.02$ at $\ell = 4$, but distance increases from 2 to 6. Figure~\ref{fig:encoding_rate} shows the encoding rate $k/n$ for base codes across all three tessellation families.

\begin{figure}[htbp]
\centering
\begin{tikzpicture}
\begin{axis}[
  width=0.85\columnwidth,
  height=0.55\columnwidth,
  xmode=log,
  ymode=log,
  xlabel={Physical qubits $n$},
  ylabel={Logical qubits $k$},
  xmin=10, xmax=1500,
  ymin=3, ymax=200,
  grid=both,
  grid style={line width=0.1pt, draw=gray!20},
  major grid style={line width=0.2pt, draw=gray!40},
  legend style={at={(0.03,0.97)}, anchor=north west, font=\small, cells={anchor=west}},
  tick label style={font=\small},
  label style={font=\small},
]
% {8,3} base codes
\addplot+[only marks, mark=o, thick, mark size=2.5pt] coordinates {
  (16,4) (32,6) (64,10) (144,20) (256,34) (336,44) (432,56)
};
\addlegendentry{$\{8,3\}$ base}
% {10,3} base codes
\addplot+[only marks, mark=square, thick, mark size=2.5pt] coordinates {
  (50,12) (120,26) (250,52) (720,146)
};
\addlegendentry{$\{10,3\}$ base}
% {12,3} base codes
\addplot+[only marks, mark=triangle, thick, mark size=2.5pt] coordinates {
  (48,14) (72,20) (96,26) (168,44) (312,80)
};
\addlegendentry{$\{12,3\}$ base}
% Asymptotic line: k = n/8 for {8,3}
\addplot[dashed, blue!60, thin, forget plot, domain=10:1500, samples=2] {x/8};
\node[font=\scriptsize, blue!60, anchor=south west] at (axis cs:800,100) {$k = n/8$};
% Asymptotic line: k = n/5 for {10,3}
\addplot[dashed, red!60, thin, forget plot, domain=10:1500, samples=2] {x/5};
\node[font=\scriptsize, red!60, anchor=south west] at (axis cs:600,120) {$k = n/5$};
% Asymptotic line: k = n/4 for {12,3}
\addplot[dashed, brown!60, thin, forget plot, domain=10:1500, samples=2] {x/4};
\node[font=\scriptsize, brown!60, anchor=south west] at (axis cs:400,150) {$k = n/4$};
\end{axis}
\end{tikzpicture}
\caption{Logical qubits $k$ versus physical qubits $n$ for base hyperbolic Floquet codes across three tessellation families. Dashed lines show asymptotic scaling: $k/n \to 1/8$ for $\{8,3\}$~\cite{higgott2024constructions}, $k/n \to 1/5$ for $\{10,3\}$, and $k/n \to 1/4$ for $\{12,3\}$.}
\label{fig:encoding_rate}
\end{figure}

%==============================================================================
\section{Noise Model Results}
%==============================================================================

\indent We evaluate our codes under the following noise models. Phenomenological noise (Section~\ref{sec:phenomenological}) provides baseline threshold estimates following prior work. Circuit-level EM3 ancilla noise (Section~\ref{sec:circuit_level}) models ancilla-based measurement circuits. The final two models target hardware with native pair measurements: SDEM3 (Section~\ref{sec:sdem3}) captures depolarizing noise motivated by Majorana tetron architectures, and the photon loss model (Section~\ref{sec:erasure}) captures heralded photon loss on spin-optical links. In the photon loss model, each loss event is treated as an erasure.

\subsection{Phenomenological Noise Results}
\label{sec:phenomenological}

\subsubsection{Phenomenological Noise Model}

\indent Following prior work on hyperbolic Floquet codes~\cite{fahimniya2023faulttolerant}, we employ a phenomenological noise model with physical error rate $p_e$:
\begin{itemize}
\item Each physical qubit independently experiences a depolarizing error with probability $p_e$ per round (X, Y, or Z each with probability $p_e/3$).
\item Each measurement outcome flips with probability $p_e$.
\end{itemize}

\subsubsection{Logical Error Rate Metrics}

\indent We report results using three logical error rate metrics, depending on the noise model:
\begin{itemize}
\item \emph{Any-logical}: A trial fails if any of the $k = 2g$ logical operators is affected by an uncorrected error. Used for phenomenological and EM3 ancilla noise.
\item \emph{Per-observable}: The average per-operator failure rate across all $k$ logical operators. Used for SDEM3 and HCF where codes have different $k$ values; also reported for phenomenological noise to enable comparison with prior work. The \emph{specific-logical} metric of Fahimniya et al.~\cite{fahimniya2023faulttolerant} is the failure rate of one particular logical operator; we use it for phenomenological threshold estimates on base codes with varying $k$.
\item \emph{Type-a/type-b classification}: For the photon loss model, each simulation instance samples an independent loss pattern (which measurements failed). Since every loss event is heralded, the decoder is given the erasure locations (aware decoding) and decodes $N$ shots against a per-instance error model. An instance is type-a if all $N$ shots are decoded correctly, type-b otherwise. The logical error rate is LER $= 0.5 \times$ (fraction of type-b instances). This methodology follows~\cite{dessertaine2024enhanced}.
\end{itemize}

\indent The any-logical metric gives higher logical error rates because it fails if any of the $k$ logical qubits is corrupted. Per-observable rates enable fair comparison across codes with different $k$. Table~\ref{tab:metrics_by_section} summarizes the metric used in each section. Cross-family comparisons at fixed physical error rate are complicated by different $k$ values.

\begin{table}[htbp]
\centering
\caption{Logical error rate metric by section.}
\label{tab:metrics_by_section}
\begin{tabular}{lll}
\toprule
Section & Noise model & Metric \\
\midrule
\ref{sec:phenomenological} & Phenomenological & Specific-logical \\
\ref{sec:circuit_level} & EM3 ancilla & Any-logical \\
\ref{sec:sdem3} & SDEM3 & Per-observable \\
\ref{sec:erasure} & Photon loss & Type-a/type-b \\
\bottomrule
\end{tabular}
\end{table}

\subsubsection{Decoding and Statistical Uncertainty}

\indent Following~\cite{fahimniya2023faulttolerant}, we decode using minimum-weight perfect matching (MWPM)~\cite{edmonds1965paths, kolmogorov2009blossom} via PyMatching~\cite{higgott2023pymatching, higgott2025sparseblossom}. Shot counts range from 3000 to 5000 per data point and are specified in each table and figure caption. For a measured logical error rate $\hat{p}_L$ over $N$ shots, the standard error is $\sigma = \sqrt{\hat{p}_L(1 - \hat{p}_L)/N}$. At 3000 shots, a measured rate of 5\% has $\sigma \approx 0.4\%$; at 50\%, $\sigma \approx 0.9\%$. Threshold ranges reported in this paper reflect the interval between the last crossing below 50\% and the first crossing above 50\% for adjacent code sizes, not a fitted confidence interval.

\subsubsection{Phenomenological Noise}

\indent Figure~\ref{fig:phenomenological_threshold} shows the any-logical error rate versus $p_e$ for the four $\{8,3\}$ base codes with $\demb \geq 4$. The H336 and H432 curves cross near $p_e \approx 0.1\%$, consistent with Fahimniya et al.~\cite{fahimniya2023faulttolerant}. For base codes, only $\{8,3\}$ exhibits threshold crossings (${\sim}0.1$--$0.2\%$, specific-logical metric); the $\{10,3\}$ codes have varying $k$ (12--146), and all $\{12,3\}$ base codes have $\demb = 2$.

\begin{figure}[htbp]
\centering
\begin{tikzpicture}
\begin{axis}[
  width=0.85\columnwidth,
  height=0.55\columnwidth,
  xmode=log,
  ymode=log,
  xlabel={Physical error rate $p_e$},
  ylabel={Any-logical error rate},
  xmin=8e-5, xmax=6e-3,
  ymin=5e-4, ymax=1.0,
  grid=both,
  grid style={line width=0.1pt, draw=gray!20},
  major grid style={line width=0.2pt, draw=gray!40},
  legend style={at={(0.03,0.97)}, anchor=north west, font=\small, cells={anchor=west}},
  tick label style={font=\small},
  label style={font=\small},
]
% H144 (n=144, k=20)
\addplot+[mark=o, thick, mark size=2pt] coordinates {
  (0.0003, 0.002) (0.0005, 0.017) (0.001, 0.054)
  (0.0015, 0.101) (0.002, 0.204) (0.003, 0.408) (0.005, 0.767)
};
\addlegendentry{H144 ($n{=}144$, $k{=}20$)}
% H256 (n=256, k=34)
\addplot+[mark=square, thick, mark size=2pt] coordinates {
  (0.0001, 0.001) (0.0003, 0.006) (0.0005, 0.019) (0.001, 0.072)
  (0.0015, 0.146) (0.002, 0.273) (0.003, 0.540) (0.005, 0.891)
};
\addlegendentry{H256 ($n{=}256$, $k{=}34$)}
% H336 (n=336, k=44)
\addplot+[mark=triangle, thick, mark size=2.5pt] coordinates {
  (0.0005, 0.007) (0.001, 0.039)
  (0.0015, 0.119) (0.002, 0.237) (0.003, 0.567) (0.005, 0.922)
};
\addlegendentry{H336 ($n{=}336$, $k{=}44$)}
% H432 (n=432, k=56)
\addplot+[mark=diamond, thick, mark size=2.5pt] coordinates {
  (0.0005, 0.007) (0.001, 0.047)
  (0.0015, 0.128) (0.002, 0.266) (0.003, 0.661) (0.005, 0.958)
};
\addlegendentry{H432 ($n{=}432$, $k{=}56$)}
% Threshold reference
\addplot[dashed, black, thin, forget plot] coordinates {(0.001, 5e-4) (0.001, 1.0)};
\node[anchor=south, font=\scriptsize, rotate=90] at (axis cs:0.001, 0.001) {$p_e \approx 0.1\%$};
\end{axis}
\end{tikzpicture}
\caption{Any-logical error rate versus physical error rate under phenomenological noise for $\{8,3\}$ base codes with $\demb \geq 4$ (3000 shots, 14 measurement rounds). Higher-$k$ codes have steeper curves above threshold; the any-logical metric compounds across more logical operators.}
\label{fig:phenomenological_threshold}
\end{figure}

\indent Fine-grained families fix $k$ and enable proper threshold crossings for all three tessellations (Table~\ref{tab:phenom_finegrained}). The $\{8,3\}$ family ($k = 4$, $f = 2$--$5$) achieves a specific-logical threshold of $1.0$--$1.2\%$ from 3 crossings; the $\{10,3\}$ family ($k = 12$, $f = 2$--$4$) achieves $1.0$--$1.3\%$ from 2 crossings; the $\{12,3\}$ family ($k = 14$, $f = 2$--$4$) achieves ${\sim}1.2\%$ from 2 crossings. All three are an order of magnitude above the ${\sim}0.1$--$0.2\%$ base-code threshold, reflecting the increased distance from fine-graining.

\begin{table}[htbp]
\centering
\small
\caption{Specific-logical error rates (\%) under phenomenological noise for fine-grained families (3000 shots, 14 rounds). Each family has fixed $k$.}
\label{tab:phenom_finegrained}
\begin{tabular}{llrrrrrrrr}
\toprule
Family & Code & $n$ & $k$ & 0.1\% & 0.3\% & 0.5\% & 1.0\% & 1.5\% \\
\midrule
\multirow{4}{*}{$\{8,3\}$}
 & H16-f2 & 64 & 4 & 0.3 & 2.3 & 7.3 & 25.2 & 38.4 \\
 & H16-f3 & 144 & 4 & 0.0 & 0.8 & 4.1 & 23.7 & 41.6 \\
 & H16-f4 & 256 & 4 & 0.0 & 0.2 & 2.1 & 23.4 & 42.9 \\
 & H16-f5 & 400 & 4 & 0.0 & 0.1 & 1.2 & 23.2 & 45.1 \\
\midrule
\multirow{3}{*}{$\{10,3\}$}
 & H50-f2 & 200 & 12 & 0.1 & 2.2 & 9.7 & 34.6 & 44.9 \\
 & H50-f3 & 450 & 12 & 0.0 & 0.3 & 3.9 & 32.2 & 46.6 \\
 & H50-f4 & 800 & 12 & 0.0 & 0.1 & 1.6 & 32.4 & 47.7 \\
\midrule
\multirow{3}{*}{$\{12,3\}$}
 & H48-f2 & 192 & 14 & 0.3 & 4.5 & 13.5 & 37.8 & 46.6 \\
 & H48-f3 & 432 & 14 & 0.0 & 1.0 & 6.8 & 36.8 & 48.2 \\
 & H48-f4 & 768 & 14 & 0.0 & 0.2 & 3.5 & 36.6 & 48.6 \\
\bottomrule
\end{tabular}
\end{table}

\subsection{Circuit-Level Simulation}
\label{sec:circuit_level}

\subsubsection{Implementation}

\indent We implement circuit-level simulations in Stim~\cite{gidney2021stim} using the EM3 ancilla (Entangling Measurement) noise model~\cite{higgott2024constructions}. This model captures correlated errors arising from ancilla-based two-qubit measurements. The EM3 ancilla model uses an ancilla qubit for each edge measurement:
\begin{itemize}
\item XX measurement: Reset ancilla, apply XCX gates from both data qubits to ancilla, measure ancilla.
\item YY measurement: Reset ancilla, apply YCX gates from both data qubits to ancilla, measure ancilla.
\item ZZ measurement: Reset ancilla, apply CX gates from both data qubits to ancilla, measure ancilla.
\end{itemize}
For physical error rate $p_e$, we apply 31 correlated \texttt{E}$(p_e/32)$ error instructions per measurement. These cover all non-trivial combinations of two-qubit Paulis $\{I,X,Y,Z\}^{\otimes 2}$ and measurement flips. Each instruction can jointly apply a Pauli error and a measurement flip on the same edge. This preserves the Pauli--flip correlation of the EM3 model~\cite{higgott2024constructions}. The measurement schedule, detector construction, and observable tracking follow~\cite{higgott2024constructions}.

\subsubsection{Results}

\indent Tables~\ref{tab:em3_ancilla_base} and~\ref{tab:em3_ancilla_fine} (Appendix~\ref{app:error_rate_tables}) show any-logical error rates under EM3 ancilla noise (3000 shots per code) for base and fine-grained codes respectively. All three families achieve thresholds of ${\sim}1.5\%$. The $k = 4$ $\{8,3\}$ fine-grained family (f2 through f5) shows an any-logical crossing near $p_e \approx 1.5\%$ (f3 vs f4). The f5 code ($n = 400$, $\demb = 7$) further suppresses logical error rates below threshold: its any-logical rate is $0.1\%$ at $p_e = 0.5\%$, compared to $0.6\%$ for f4. The $k = 12$ $\{10,3\}$ family (f2, f3, f4) shows f4 ($n = 800$, $\demb = 8$) outperforming f3 ($n = 450$, $\demb = 6$) at all rates below ${\sim}1.5\%$. The $k = 14$ $\{12,3\}$ family (f2, f3, f4) shows a crossing near $p_e \approx 1.5\%$ (f3 vs f4). Figure~\ref{fig:em3_ancilla_threshold} shows the per-observable threshold crossing for the $k = 4$ family.

\begin{figure}[htbp]
\centering
\begin{tikzpicture}
\begin{axis}[
  width=0.85\columnwidth,
  height=0.55\columnwidth,
  xmode=log,
  ymode=log,
  xlabel={Physical error rate $p_e$},
  ylabel={Per-observable logical error rate},
  xmin=2e-3, xmax=3.5e-2,
  ymin=5e-4, ymax=0.55,
  grid=both,
  grid style={line width=0.1pt, draw=gray!20},
  major grid style={line width=0.2pt, draw=gray!40},
  legend style={at={(0.03,0.97)}, anchor=north west, font=\small, cells={anchor=west}},
  tick label style={font=\small},
  label style={font=\small},
]
% f2 (n=64, d=3)
\addplot+[mark=o, thick, mark size=2pt] coordinates {
  (0.003, 0.02033) (0.005, 0.05792) (0.007, 0.12283)
  (0.010, 0.22225) (0.012, 0.30225) (0.015, 0.39083)
  (0.020, 0.46467) (0.025, 0.49758) (0.030, 0.49892)
};
\addlegendentry{$\ell{=}2$\; ($n{=}64$, $\demb{=}3$)}
% f3 (n=144, d=4)
\addplot+[mark=square, thick, mark size=2pt] coordinates {
  (0.003, 0.00083) (0.005, 0.00883) (0.007, 0.02767)
  (0.010, 0.09000) (0.012, 0.15483) (0.015, 0.27300)
  (0.020, 0.44558) (0.025, 0.49417) (0.030, 0.49350)
};
\addlegendentry{$\ell{=}3$\; ($n{=}144$, $\demb{=}4$)}
% f4 (n=256, d=6)
\addplot+[mark=triangle, thick, mark size=2.5pt] coordinates {
  (0.005, 0.00175) (0.007, 0.00733)
  (0.010, 0.03417) (0.012, 0.08892) (0.015, 0.21467)
  (0.020, 0.42650) (0.025, 0.50058) (0.030, 0.50733)
};
\addlegendentry{$\ell{=}4$\; ($n{=}256$, $\demb{=}6$)}
% f5 (n=400, d=7)
\addplot+[mark=diamond, thick, mark size=2.5pt] coordinates {
  (0.007, 0.00133)
  (0.010, 0.01317) (0.012, 0.05125) (0.015, 0.16325)
  (0.020, 0.43017) (0.025, 0.49183) (0.030, 0.50592)
};
\addlegendentry{$\ell{=}5$\; ($n{=}400$, $\demb{=}7$)}
% Threshold reference
\addplot[dashed, black, thin, forget plot] coordinates {(0.016, 5e-4) (0.016, 0.55)};
\node[anchor=south, font=\scriptsize, rotate=90] at (axis cs:0.016, 0.001) {$p_e \approx 1.6\%$};
\end{axis}
\end{tikzpicture}
\caption{Per-observable logical error rate versus physical error rate under correlated EM3 ancilla noise for the Bolza $k{=}4$ semi-hyperbolic family (3000 shots, 32 XYZ cycles). The curves cross near $p_e \approx 1.5$--$2.0\%$. This matches the threshold reported by Higgott and Breuckmann~\cite{higgott2024constructions}.}
\label{fig:em3_ancilla_threshold}
\end{figure}

\subsection{Native Pair-Measurement Noise Model}
\label{sec:sdem3}

\subsubsection{Model Definition}

\indent SDEM3 (Single-step Depolarizing EM3)~\cite{higgott2024constructions} applies the following independent noise channels per measurement step, parameterized by physical error rate $p_e$:

\begin{itemize}
\item Two-qubit depolarizing noise with strength $15p_e/16$ before each MPP. The factor $15/16$ matches the marginal non-trivial Pauli probability: 15 non-trivial outcomes out of 16 Pauli combinations.
\item Measurement outcome flip with probability $p_e/2$ per edge.
\item Single-qubit $X$ errors with probability $p_e/2$ after initialization and before final measurement.
\item No idle noise for base codes (every qubit participates in each round on the 3-valent graph). Fine-grained codes introduce subdivision vertices with valence up to 6, so some qubits are idle in certain rounds. Our simulations omit idle noise on these qubits.
\end{itemize}

\indent This decomposition reproduces the marginal error rates of the EM3 noise model~\cite{higgott2024constructions}. SDEM3 treats Pauli errors and flips independently, while EM3 correlates them within each outcome. This correlation loss affects thresholds: SDEM3 achieves ${\sim}1.0$--$1.2\%$ for all three families, while the correlated EM3 ancilla model (Section~\ref{sec:circuit_level}) achieves ${\sim}1.5\%$.

\subsubsection{Monolithic SDEM3 Results}

\indent Table~\ref{tab:sdem3_monolithic} (Appendix~\ref{app:error_rate_tables}) presents per-observable error rates for same-$k$ fine-grained families under SDEM3 noise. The Bolza $k{=}4$ family includes five fine-graining levels ($\ell = 2$ through $6$); the other families have three levels each. All families show per-observable threshold crossings near $p_e \approx 1.0$--$1.2\%$: within each family, the larger code achieves lower error rates at every $p_e$ below this threshold, and all codes saturate to ${\sim}50\%$ above it. At $p_e = 0.5\%$, the largest codes in each family achieve per-observable error rates of 0.4\% ($\{8,3\}$ $k{=}4$, f6), 1.1\% ($\{10,3\}$ $k{=}12$, f4), and 2.2\% ($\{12,3\}$ $k{=}14$, f4). Figure~\ref{fig:sdem3_threshold} shows the per-observable threshold crossing for the $k = 4$ family under SDEM3.

\begin{figure}[htbp]
\centering
\begin{tikzpicture}
\begin{axis}[
  width=0.85\columnwidth,
  height=0.55\columnwidth,
  xmode=log,
  ymode=log,
  xlabel={Physical error rate $p_e$},
  ylabel={Per-observable logical error rate},
  xmin=1e-3, xmax=1.6e-2,
  ymin=1e-3, ymax=0.55,
  grid=both,
  grid style={line width=0.1pt, draw=gray!20},
  major grid style={line width=0.2pt, draw=gray!40},
  legend style={at={(0.03,0.97)}, anchor=north west, font=\small, cells={anchor=west}},
  tick label style={font=\small},
  label style={font=\small},
  cycle list name=color list,
]
% l=2 (n=64, d=3)
\addplot+[mark=o, thick, mark size=2pt] coordinates {
  (0.001, 0.006) (0.002, 0.023) (0.003, 0.062) (0.005, 0.17)
  (0.007, 0.292) (0.01, 0.432) (0.012, 0.473) (0.015, 0.493)
};
\addlegendentry{$\ell{=}2$\; ($n{=}64$, $\demb{=}3$)}
% l=3 (n=144, d=4)
\addplot+[mark=square, thick, mark size=2pt] coordinates {
  (0.002, 0.002) (0.003, 0.009) (0.005, 0.057)
  (0.007, 0.175) (0.01, 0.381) (0.012, 0.46) (0.015, 0.496)
};
\addlegendentry{$\ell{=}3$\; ($n{=}144$, $\demb{=}4$)}
% l=4 (n=256, d=6)
\addplot+[mark=triangle, thick, mark size=2.5pt] coordinates {
  (0.003, 0.002) (0.005, 0.019)
  (0.007, 0.102) (0.01, 0.342) (0.012, 0.452) (0.015, 0.498)
};
\addlegendentry{$\ell{=}4$\; ($n{=}256$, $\demb{=}6$)}
% l=5 (n=400, d=7)
\addplot+[mark=diamond, thick, mark size=2.5pt] coordinates {
  (0.005, 0.007) (0.007, 0.057) (0.01, 0.324) (0.012, 0.462) (0.015, 0.5)
};
\addlegendentry{$\ell{=}5$\; ($n{=}400$, $\demb{=}7$)}
% l=6 (n=576, d=8)
\addplot+[mark=pentagon, thick, mark size=2.5pt] coordinates {
  (0.005, 0.004) (0.007, 0.038) (0.01, 0.31) (0.012, 0.463) (0.015, 0.5)
};
\addlegendentry{$\ell{=}6$\; ($n{=}576$, $\demb{=}8$)}
% Threshold reference line
\addplot[dashed, black, thin, forget plot] coordinates {(0.011, 1e-3) (0.011, 0.55)};
\node[anchor=south, font=\scriptsize, rotate=90] at (axis cs:0.011, 0.002) {$p_e \approx 1.1\%$};
\end{axis}
\end{tikzpicture}
\caption{Per-observable logical error rate versus physical error rate under SDEM3 noise for the Bolza $k{=}4$ semi-hyperbolic family (5000 shots, 32 XYZ cycles). Fine-graining levels $\ell = 2$ through $6$ correspond to $n = 64$ to $576$ qubits and embedded distances $\demb = 3$ to $8$. The curves cross near $p_e \approx 1.0$--$1.2\%$.}
\label{fig:sdem3_threshold}
\end{figure}

\subsection{Photon Loss Noise Model}
\label{sec:erasure}

\indent Spin-optical hardware performs native pair measurements via photonic entanglement. The dominant error mechanism is photon loss. In the MZZ-SPOQC architecture, every loss event is heralded: the platform detects which measurements failed, and the decoder treats these failed measurements as erasures. Loss and erasure are distinct concepts in general---loss is the physical process, while erasure is the information-theoretic consequence that the decoder knows which locations were corrupted---but on MZZ-SPOQC they coincide because every loss event is detected~\cite{dessertaine2024enhanced}. Under this noise model, the honeycomb Floquet code achieves a photon loss threshold of ${\sim}6.3\%$, roughly twice the ${\sim}3.3\%$ surface code threshold~\cite{dessertaine2024enhanced}. We apply the same noise model and decoding strategy to our hyperbolic Floquet codes to determine whether hyperbolic geometry raises the photon loss threshold further.

\subsubsection{Noise Model}

\indent We adopt the photon loss model of~\cite{dessertaine2024enhanced}. Each pair measurement independently fails with probability $p_{\text{RUS}}$ due to a failed repeat-until-success (RUS) protocol. A failed measurement is heralded and treated as an erasure by the decoder. The single-photon loss rate $\varepsilon$ and the RUS failure probability are related by
\begin{equation}
\label{eq:p_rus}
p_{\text{RUS}} = \frac{2 - 2(1-\varepsilon)^2}{2 - (1-\varepsilon)^2}.
\end{equation}
This relationship arises from the SPOQC-2 architecture~\cite{dessertaine2024enhanced}: each two-qubit measurement requires two photons, and the protocol succeeds only if both photons are detected. Let $s = (1-\varepsilon)^2$ be the probability that both photons survive. Each RUS attempt succeeds with probability $s$. See~\cite{dessertaine2024enhanced}, Section III.B for details.

\subsubsection{Simulation Methodology}

\indent Each simulation instance proceeds as follows. We generate a noiseless hyperbolic Floquet circuit, then sample an independent loss pattern: each pair measurement fails (is lost) with probability $p_{\text{RUS}}$. Since every loss event is heralded, the decoder knows which measurements were erased. We inject the corresponding dephasing errors at erased locations and construct a detector error model (DEM) specific to that erasure pattern. PyMatching MWPM decodes against this per-instance DEM (aware decoding). We repeat for $N$ shots per instance and classify each instance as type-a (all shots correct) or type-b (at least one shot fails). The logical error rate is
\begin{equation}
\text{LER} = \tfrac{1}{2} \times (\text{fraction of type-b instances}),
\end{equation}
as in~\cite{dessertaine2024enhanced}.

\subsubsection{Hyperbolic Erasure Thresholds}

\indent Tables~\ref{tab:erasure_hyp} and~\ref{tab:erasure_all} (Appendix~\ref{app:error_rate_tables}) report photon loss LER across fine-grained codes. The Bolza family ($k = 4$, fine-grained levels $\ell = 2$ through $5$) achieves a photon loss threshold of $\varepsilon \approx 8.5$--$9\%$. Among the three tessellation families, the $\{8,3\}$ Bolza family shows the widest threshold window. Figure~\ref{fig:erasure_threshold} shows the threshold crossing for this family along with the planar honeycomb reference.

\begin{figure}[htbp]
\centering
\begin{tikzpicture}
\begin{axis}[
  width=0.85\columnwidth,
  height=0.55\columnwidth,
  xmode=log,
  ymode=log,
  xlabel={Photon loss rate $\varepsilon$},
  ylabel={Logical error rate (type-a/b)},
  xmin=2e-2, xmax=0.1,
  ymin=1e-3, ymax=0.55,
  grid=both,
  grid style={line width=0.1pt, draw=gray!20},
  major grid style={line width=0.2pt, draw=gray!40},
  legend style={at={(0.03,0.97)}, anchor=north west, font=\small, cells={anchor=west}},
  tick label style={font=\small},
  label style={font=\small},
]
% f2 (n=64, d=3)
\addplot+[mark=o, thick, mark size=2pt] coordinates {
  (0.02667, 0.004) (0.04140, 0.032) (0.05074, 0.096)
  (0.05719, 0.188) (0.06384, 0.264)
  (0.07418, 0.408) (0.08501, 0.482) (0.09251, 0.496)
};
\addlegendentry{$\ell{=}2$\; ($n{=}64$, $\demb{=}3$)}
% f3 (n=144, d=4)
\addplot+[mark=square, thick, mark size=2pt] coordinates {
  (0.04140, 0.004) (0.05074, 0.016) (0.05719, 0.048)
  (0.06384, 0.128) (0.07418, 0.280)
  (0.08501, 0.452) (0.09251, 0.490)
};
\addlegendentry{$\ell{=}3$\; ($n{=}144$, $\demb{=}4$)}
% f4 (n=256, d=6)
\addplot+[mark=triangle, thick, mark size=2.5pt] coordinates {
  (0.05074, 0.004) (0.05719, 0.008) (0.06384, 0.042)
  (0.07418, 0.200) (0.08501, 0.432) (0.09251, 0.496)
};
\addlegendentry{$\ell{=}4$\; ($n{=}256$, $\demb{=}6$)}
% f5 (n=400, d=7)
\addplot+[mark=diamond, thick, mark size=2.5pt] coordinates {
  (0.05074, 0.002) (0.05719, 0.006) (0.06384, 0.030)
  (0.07418, 0.154) (0.08501, 0.426) (0.09251, 0.486)
};
\addlegendentry{$\ell{=}5$\; ($n{=}400$, $\demb{=}7$)}
% Honeycomb threshold reference
\addplot[dashed, gray, thin, forget plot] coordinates {(0.063, 1e-3) (0.063, 0.55)};
\node[anchor=south, font=\scriptsize, gray, rotate=90] at (axis cs:0.063, 0.002) {honeycomb $6.3\%$};
% Hyperbolic threshold reference
\addplot[dashed, black, thin, forget plot] coordinates {(0.088, 1e-3) (0.088, 0.55)};
\node[anchor=south, font=\scriptsize, rotate=90] at (axis cs:0.088, 0.002) {hyperbolic ${\sim}8.5$--$9\%$};
\end{axis}
\end{tikzpicture}
\caption{Logical error rate versus photon loss rate $\varepsilon$ under the SPOQC-2 photon loss model for the Bolza $k{=}4$ family ($M = 250$ instances, $N = 250$ shots, 8 XYZ cycles). Each loss event is heralded and modeled as an erasure. The dashed gray line marks the planar honeycomb photon loss threshold of $6.3\%$~\cite{dessertaine2024enhanced}. The hyperbolic curves converge near $\varepsilon \approx 8.5$--$9\%$.}
\label{fig:erasure_threshold}
\end{figure}

\subsubsection{Fault-Tolerant Region Area}

\indent Dessertaine et al.~\cite{dessertaine2024enhanced} quantify fault tolerance under the SPOQC-2 noise model by sweeping multiple noise parameters simultaneously and measuring the region where logical error rate decreases with increasing code distance. We apply their radial threshold extraction methodology to compare planar and hyperbolic codes on a like-for-like basis.

\indent We calibrated our SPOQC-2 implementation against Dessertaine's honeycomb 2D fault-tolerant area benchmark. The calibration uses $d=17$ versus $d=19$, radial threshold extraction, and the single-logical observable mismatch metric from the planar circuits. For each radial direction, we select the best threshold across tested $T_{\max}$ values and discard saturated crossings where both code distances are near $50\%$ logical error rate (min-LER $\ge 45\%$). With this protocol, the honeycomb area is $7.29\times 10^{-5}$, close to the reported $7.7\times 10^{-5}$.

\indent We ran the same radial post-processing on the fixed-$k$ H64 fine-grained pair (H64$_{8,3}$-f3 versus H64$_{8,3}$-f4, both $k=10$) using the per-observable metric. With the same $T_{\max}$ envelope and the same non-saturated crossing filter, the extracted 2D area is $4.81\times 10^{-5}$. This is $0.62\times$ the honeycomb area ($7.7\times 10^{-5}$) and $2.2\times$ the surface code compiled to pair measurements ($2.2\times 10^{-5}$, MZZ-SPOQC pentagon tiling~\cite{dessertaine2024enhanced}). The hyperbolic codes encode $k = 10$ logical qubits versus $k = 1$ for the surface code.

\subsection{Summary and Comparison}

\indent Table~\ref{tab:comprehensive_summary} consolidates threshold results across the three circuit-level noise models, using same-$k$ fine-grained code crossings with three or more codes per family.

\begin{table}[htbp]
\centering
\caption{Threshold summary across noise models for fine-grained codes (same-$k$ crossings with three or more codes per family). Metric: EM3 ancilla = any-logical, SDEM3 = per-observable, erasure = type-a/type-b (see Table~\ref{tab:metrics_by_section}). Erasure threshold is in $\varepsilon$ (\%).}
\label{tab:comprehensive_summary}
\begin{tabular}{lcccc}
\toprule
Family & EM3 ancilla & SDEM3 & Erasure \\
\midrule
$\{8,3\}$ & ${\sim}1.5\%$ & 1.0--1.2\% & 8.5--9\% \\
$\{10,3\}$ & ${\sim}1.5\%$ & 1.0--1.2\% & 7--8\% \\
$\{12,3\}$ & ${\sim}1.5\%$ & 1.0--1.2\% & 6.5--8\% \\
\bottomrule
\end{tabular}
\end{table}

\begin{table}[htbp]
\centering
\caption{Comparison with prior work on hyperbolic Floquet codes. Thresholds marked $^\dagger$ are base-code thresholds; all others use same-$k$ fine-grained families with three or more codes.}
\label{tab:literature_comparison}
\begin{tabular}{llcc}
\toprule
Reference & Noise model & Threshold & $k$ \\
\midrule
Higgott \& Breuckmann~\cite{higgott2024constructions} ($\{8,3\}$) & EM3 (correlated) & ${\sim}1.5$--$2\%$ & 4--674 \\
Fahimniya et al.~\cite{fahimniya2023faulttolerant} ($\{8,3\}$) & Phenomenological & ${\sim}0.1\%^\dagger$ & 4--272 \\
\midrule
$\{8,3\}$ (this work) & Phenomenological (base, specific-logical) & ${\sim}0.1$--$0.2\%^{\dagger}$ & 6--56 \\
$\{8,3\}$ (this work) & Phenomenological (fine-grained) & $1.0$--$1.2\%$ & 4 \\
$\{8,3\}$ (this work) & EM3 ancilla & ${\sim}1.5\%$ & 4 \\
$\{8,3\}$ (this work) & SDEM3 & $1.0$--$1.2\%$ & 4 \\
$\{8,3\}$ (this work) & Erasure & $8.5$--$9\%$ & 4 \\
\midrule
$\{10,3\}$ (this work) & Phenomenological (fine-grained) & $1.0$--$1.3\%$ & 12 \\
$\{10,3\}$ (this work) & EM3 ancilla & ${\sim}1.5\%$ & 12 \\
$\{10,3\}$ (this work) & SDEM3 & $1.0$--$1.2\%$ & 12 \\
$\{10,3\}$ (this work) & Erasure & $7$--$8\%$ & 12 \\
\midrule
$\{12,3\}$ (this work) & Phenomenological (fine-grained) & ${\sim}1.2\%$ & 14 \\
$\{12,3\}$ (this work) & EM3 ancilla & ${\sim}1.5\%$ & 14 \\
$\{12,3\}$ (this work) & SDEM3 & $1.0$--$1.2\%$ & 14 \\
$\{12,3\}$ (this work) & Erasure & $6.5$--$8\%$ & 14 \\
\bottomrule
\end{tabular}
\end{table}

\indent Table~\ref{tab:literature_comparison} compares our results with prior work. Our EM3 ancilla threshold of ${\sim}1.5\%$ is consistent with the ${\sim}1.5$--$2\%$ reported by Higgott and Breuckmann~\cite{higgott2024constructions} using the same correlated noise model on semi-hyperbolic Bolza ($k = 4$) codes. Our SDEM3 thresholds of $1.0$--$1.2\%$ for all three families are lower than the correlated EM3 values because SDEM3 loses the Pauli--flip correlation. All thresholds in Table~\ref{tab:comprehensive_summary} are computed by comparing per-observable error rates across codes with the same $k$ within each family ($\ell = 2, 3, 4$ fine-graining levels). Distributed noise models are evaluated in a companion paper~\cite{galimova2026distributed}.

\begin{table}[htbp]
\centering
\caption{Photon loss threshold comparison under the SPOQC-2 model. Each loss event is heralded and modeled as an erasure. Surface code thresholds from Dessertaine et al.~\cite{dessertaine2024enhanced}: CZ-SPOQC uses auxiliary-qubit syndrome extraction; MZZ-SPOQC uses the pentagon tiling pair-measurement circuit.}
\label{tab:erasure_comparison}
\begin{tabular}{@{}lcc@{}}
\toprule
Code family & Threshold & $k$ \\
\midrule
Surface, CZ~\cite{dessertaine2024enhanced} & ${\sim}3.3\%$ & 1 \\
Surface, MZZ~\cite{dessertaine2024enhanced} & ${\sim}4.3\%$ & 1 \\
Honeycomb, MZZ~\cite{dessertaine2024enhanced} & ${\sim}6.3\%$ & 1 \\
\midrule
$\{8,3\}$, MZZ (this work) & ${\sim}8.5$--$9\%$ & 4 \\
$\{10,3\}$, MZZ (this work) & ${\sim}7$--$8\%$ & 12 \\
$\{12,3\}$, MZZ (this work) & ${\sim}6.5$--$8\%$ & 14 \\
\bottomrule
\end{tabular}
\end{table}

%==============================================================================
\section{Conclusion}
%==============================================================================

\indent The three tessellation families exhibit a tradeoff between encoding rate and error correction. The $\{8,3\}$ family has the lowest genus-per-vertex ratio ($g/V \to 6.25\%$) and the best thresholds across all noise models. The $\{12,3\}$ family achieves the highest encoding rate ($k/n \to 25\%$) but all base codes have $\demb = 2$, requiring fine-graining for error correction. The $\{10,3\}$ family falls between the two on both measures.

\indent All three families achieve comparable depolarizing thresholds (${\sim}1.5\%$ under EM3, ${\sim}1.0$--$1.2\%$ with SDEM3). The families separate for photon loss: the $\{8,3\}$ family reaches ${\sim}8.5$--$9\%$, exceeding both the planar honeycomb threshold of ${\sim}6.3\%$~\cite{dessertaine2024enhanced} and the surface code threshold of ${\sim}3.3\%$~\cite{dessertaine2024enhanced}. In the multi-parameter SPOQC-2 noise model, the $\{8,3\}$ codes achieve a 2D fault-tolerant area $2.2\times$ that of the surface code compiled to pair measurements.

\indent These results suggest that platforms with native pair measurements---Majorana tetrons and spin-optical architectures---are natural targets for hyperbolic Floquet codes. The codes require only weight-2 measurements but demand all-to-all connectivity within each face of the tessellation. This connectivity requirement is compatible with modular architectures where each module contains one face. In a companion paper~\cite{galimova2026distributed}, we evaluate the same code families in a distributed setting with heterogeneous local and non-local noise.

\subsection{Future Directions}

\begin{enumerate}
\item Decoder optimization: our simulations use uniform edge weights in MWPM. Optimizing weights based on error probabilities and code structure could raise thresholds.
\item Logical gates: this work addresses quantum memory only. Lattice surgery and Dehn twists have been proposed for hyperbolic surface codes~\cite{lavasani2019universal}.
\item Near-term demonstrations: hardware with native two-qubit parity measurements could implement these codes directly.
\item HCF schedule: the Hyperbolic Color Floquet (HCF) schedule~\cite{ozawa2025hyperbolic} uses only XX and ZZ measurements (no YY). This produces graph-edge syndromes amenable to standard MWPM decoding. The HCF schedule cycles through six steps: XX on red, ZZ on green, XX on blue, ZZ on red, XX on green, ZZ on blue. We could evaluate additional noise models for hyperbolic Floquet codes with the HCF schedule.
\end{enumerate}

%==============================================================================
\section*{Acknowledgments}

We thank Th\'eo Dessertaine for discussions.

%==============================================================================

%==============================================================================
\appendix
\section{Error Rate Tables}
\label{app:error_rate_tables}

\indent This appendix collects the per-code logical error rate data underlying the threshold estimates in Table~\ref{tab:comprehensive_summary}.

\begin{table}[H]
\centering
\caption{Any-logical error rates (\%) for $\demb \geq 4$ base hyperbolic codes under EM3\_ancilla noise (3000 shots). All $\{12,3\}$ base codes have $\demb = 2$ and are excluded.}
\label{tab:em3_ancilla_base}
\resizebox{\columnwidth}{!}{%
\begin{tabular}{lrrrrrrrrrrr}
\toprule
Code & $n$ & $k$ & $\demb$ & 0.01\% & 0.03\% & 0.05\% & 0.1\% & 0.2\% & 0.3\% & 0.5\% & 1\% \\
\midrule
H144$_{8,3}$ & 144 & 20 & 4 & 0.0 & 0.1 & 0.9 & 2.5 & 10.8 & 26.1 & 65.8 & 99.7 \\
H256$_{8,3}$ & 256 & 34 & 4 & 0.0 & 0.4 & 1.1 & 3.1 & 14.7 & 32.2 & 75.6 & 100.0 \\
H336$_{8,3}$ & 336 & 44 & 4 & 0.0 & 0.1 & 0.1 & 1.3 & 8.6 & 21.8 & 70.1 & 100.0 \\
H432$_{8,3}$ & 432 & 56 & 4 & 0.0 & 0.1 & 0.4 & 1.8 & 10.3 & 27.4 & 77.8 & 100.0 \\
\midrule
H250$_{10,3}$ & 250 & 52 & 4 & 0.0 & 1.2 & 4.1 & 14.2 & 46.2 & 78.3 & 99.1 & 100.0 \\
\bottomrule
\end{tabular}}
\end{table}

\begin{table}[H]
\centering
\caption{Any-logical error rates (\%) for same-$k$ fine-grained families with three or more codes under EM3\_ancilla noise (3000 shots). The column $\ell$ denotes fine-graining level.}
\label{tab:em3_ancilla_fine}
\begin{tabular}{lrrrrrrrrr}
\toprule
Code & $n$ & $k$ & $\demb$ & $\ell$ & 0.1\% & 0.2\% & 0.3\% & 0.5\% & 1\% \\
\midrule
H16$_{8,3}$-f2 & 64 & 4 & 3 & 2 & 0.6 & 2.5 & 5.7 & 16.6 & 57.2 \\
H16$_{8,3}$-f3 & 144 & 4 & 4 & 3 & 0.0 & 0.2 & 0.6 & 3.1 & 33.2 \\
H16$_{8,3}$-f4 & 256 & 4 & 6 & 4 & 0.0 & 0.0 & 0.2 & 0.6 & 21.1 \\
H16$_{8,3}$-f5 & 400 & 4 & 7 & 5 & 0.0 & 0.0 & 0.0 & 0.1 & 11.6 \\
\midrule
H50$_{10,3}$-f2 & 200 & 12 & 4 & 2 & 0.1 & 0.8 & 3.6 & 16.3 & 84.5 \\
H50$_{10,3}$-f3 & 450 & 12 & 6 & 3 & 0.0 & 0.1 & 0.3 & 2.4 & 55.1 \\
H50$_{10,3}$-f4 & 800 & 12 & 8 & 4 & 0.0 & 0.0 & 0.0 & 0.3 & 28.7 \\
\midrule
H48$_{12,3}$-f2 & 192 & 14 & 4 & 2 & 1.6 & 7.0 & 16.6 & 47.6 & 97.2 \\
H48$_{12,3}$-f3 & 432 & 14 & 4 & 3 & 0.0 & 0.4 & 1.2 & 7.1 & 68.3 \\
H48$_{12,3}$-f4 & 768 & 14 & 7 & 4 & 0.0 & 0.1 & 0.1 & 1.9 & 51.3 \\
\bottomrule
\end{tabular}
\end{table}

\begin{table}[H]
\centering
\caption{Per-observable logical error rate (\%) under SDEM3 noise for the Bolza $k{=}4$ semi-hyperbolic family and same-$k$ fine-grained families from the three tessellations (5000 shots). Within each family, the larger code achieves lower error rates at all $p_e$ below the per-observable threshold of ${\approx}1.0$--$1.2\%$.}
\label{tab:sdem3_monolithic}
\begin{tabular}{llrrrrrrr}
\toprule
Family & Code & $n$ & $k$ & 0.3\% & 0.5\% & 0.7\% & 1.0\% & 1.2\% \\
\midrule
\multirow{5}{*}{$\{8,3\}$}
& H16$_{8,3}$-f2 & 64 & 4 & 6.4 & 16.9 & 28.8 & 42.3 & 46.7 \\
& H16$_{8,3}$-f3 & 144 & 4 & 1.0 & 5.4 & 17.1 & 38.3 & 46.0 \\
& H16$_{8,3}$-f4 & 256 & 4 & 0.2 & 2.4 & 10.1 & 34.3 & 46.1 \\
& H16$_{8,3}$-f5 & 400 & 4 & 0.0 & 0.7 & 5.7 & 32.4 & 46.2 \\
& H16$_{8,3}$-f6 & 576 & 4 & 0.0 & 0.4 & 3.8 & 31.0 & 46.3 \\
\midrule
\multirow{3}{*}{$\{10,3\}$}
& H50$_{10,3}$-f2 & 200 & 12 & 3.1 & 14.6 & 32.5 & 47.2 & 49.3 \\
& H50$_{10,3}$-f3 & 450 & 12 & 0.3 & 3.6 & 18.2 & 45.1 & 49.4 \\
& H50$_{10,3}$-f4 & 800 & 12 & 0.0 & 1.1 & 10.4 & 43.1 & 49.2 \\
\midrule
\multirow{3}{*}{$\{12,3\}$}
& H48$_{12,3}$-f2 & 192 & 14 & 7.4 & 23.7 & 40.7 & 49.1 & 49.9 \\
& H48$_{12,3}$-f3 & 432 & 14 & 1.1 & 7.7 & 26.7 & 48.1 & 50.0 \\
& H48$_{12,3}$-f4 & 768 & 14 & 0.2 & 2.2 & 16.0 & 46.2 & 49.8 \\
\bottomrule
\end{tabular}
\end{table}

\begin{table}[H]
\centering
\caption{Erasure LER (\%) for hyperbolic Floquet codes, Bolza family ($k = 4$). f2--f3: $M = 250$, $N = 250$; f4--f5: $M = 200$, $N = 64$. Larger codes outperform at all $\varepsilon$ below ${\sim}8.5\%$.}
\label{tab:erasure_hyp}
\resizebox{\columnwidth}{!}{%
\begin{tabular}{lrrccccccccc}
\toprule
Code & $n$ & $\demb$ & 2.7\% & 4.1\% & 5.1\% & 5.7\% & 6.4\% & 7.4\% & 8.5\% & 9.3\% \\
\midrule
f2 & 64 & 3 & 0.4 & 3.2 & 9.6 & 18.8 & 26.4 & 40.8 & 48.2 & 49.6 \\
f3 & 144 & 4 & 0.0 & 0.4 & 1.6 & 4.8 & 12.8 & 28.0 & 45.2 & 49.0 \\
f4 & 256 & 6 & 0.0 & 0.0 & --- & 1.0 & 4.5 & 20.5 & 43.0 & 49.5 \\
f5 & 400 & 7 & 0.0 & 0.0 & --- & 0.8 & 3.0 & 15.8 & 41.8 & 48.2 \\
\bottomrule
\end{tabular}}
\end{table}

\begin{table}[H]
\centering
\caption{Erasure LER (\%) for same-$k$ fine-grained families with three or more codes. Sampling ranges across all rows are $M = 80$--$500$ instances and $N = 32$--$500$ shots per instance.}
\label{tab:erasure_all}
\begin{tabular}{llrrcccccc}
\toprule
Family & Code & $n$ & $k$ & 4.1\% & 5.7\% & 6.4\% & 7.4\% & 8.5\% \\
\midrule
\multirow{3}{*}{$\{8,3\}$}
& f2 ($k{=}4$)  & 64 & 4  & 2.7 & 16.7 & 26.4 & 39.8 & 47.6 \\
& f3 ($k{=}4$)  & 144 & 4 & 0.4 & 4.8  & 12.8 & 28.0 & 45.2 \\
& f4 ($k{=}4$)  & 256 & 4 & 0.0 & 1.0  & 4.5  & 20.5 & 43.0 \\
\midrule
\multirow{3}{*}{$\{10,3\}$}
& f2 ($k{=}12$) & 200 & 12 & 2.8 & 21.8 & 37.8 & 49.4 & 49.8 \\
& f3 ($k{=}12$) & 450 & 12 & 0.0 & 4.6  & 14.2 & 44.2 & 50.0 \\
& f4 ($k{=}12$) & 800 & 12 & 0.0 & 0.6  & 5.0  & 33.1 & 50.0 \\
\midrule
\multirow{3}{*}{$\{12,3\}$}
& f2 ($k{=}14$) & 192 & 14 & 11.2 & 39.4 & 46.2 & 50.0 & 50.0 \\
& f3 ($k{=}14$) & 432 & 14 & 0.6  & 15.6 & 32.5 & 47.5 & 50.0 \\
& f4 ($k{=}14$) & 768 & 14 & 0.0  & 3.1  & 11.9 & 40.6 & 50.0 \\
\bottomrule
\end{tabular}
\end{table}

\end{document}